\documentclass{PoS}

\usepackage{graphicx}
\usepackage{natbib}
\usepackage{txfonts}
\usepackage{epsfig}

\title{LOFT, the Large Observatory For X-ray Timing}

\ShortTitle{LOFT, the Large Observatory For X-ray Timing}

\author{\speaker{E. Bozzo}\\
        ISDC Data Centre for Astrophysics, Chemin d'Ecogia 16,
             CH-1290 Versoix, Switzerland;\\
        E-mail: \email{enrico.bozzo@unige.ch}}

\author{J. W. den Herder\\
        SRON, the Netherlands Institute for Space Research, Sorbonnelaan 2, 3584 CA Utrecht, the Netherlands \\
        }
        
 \author{M. Feroci\\
        INAF/IASF, Via del Fosso del Cavaliere 100, I-00133 Roma, Italy\\
        }  
        
  \author{L. Stella\\
        INAF-OAR, Via di Frascati 33, I-00040 Monteporzio Catone, Italy \\
        }         
        
  \author{on the behalf of the LOFT consortium\\
        }            

\abstract{The Large Observatory For X-ray Timing, LOFT, was selected by the European Space Agency as one of the four 
Cosmic Vision M3 candidate missions to compete for a launch opportunity at the start of 
the 2020s. Thanks to an innovative design and the development of large-area monolithic silicon
drift detectors, the Large Area Detector (LAD) on board LOFT will operate in the 2-30 keV range 
(up to 50 keV in expanded mode), and achieve an effective area of $\sim$10~m$^2$ at 8~keV, a time resolution of $\sim$10~$\mu$s, and  
a spectral resolution of $\sim$260 eV (FWHM at 6~keV). These characteristics make LOFT a perfectly suited instrument to perform 
high-time-resolution X-ray observations of collapsed objects in our galaxy
and brightest supermassive black holes in active galactic nuclei. 
LOFT will yield unprecedented information on strongly curved spacetimes and matter under
extreme conditions of pressure and magnetic field strength, thus addressing two of the fundamental 
questions of the Cosmic Vision Theme ``Matter under extreme conditions'': does matter orbiting close to the event horizon follow the
predictions of general relativity? What is the equation of state of matter in
neutron stars? 
}

\FullConference{The Extreme and Variable High Energy Sky - extremesky2011,\\
		September 19-23, 2011\\
		Chia Laguna (Cagliari), Italy}

\begin{document}

\section{Introduction}

Collapsed objects, like neutron stars (NS) and black holes (BH), are characterized by exceptionally intense gravitational and magnetic 
fields. They provide a unique opportunity to reveal a variety of general relativistic effects and probe the conditions of matter 
in presence of supercritical magnetic fields and bulk densities exceeding those of atomic nuclei \citep{psaltis08}. Carrying out these investigations 
requires the exploitation of diagnostics that are capable of probing {\it in situ} matter motion in close vicinity of collapsed objects 
and/or measuring the rotation of the collapsed object itself. 
In this respect, both techniques based on X-ray spectroscopy and fast X-ray timing measurements are crucial. 
The former permit to reveal very broad profiles of X-ray emission lines (Fe K-shell in
particular) from accretion disks extending down to the radius of the NSs and BHs marginally stable orbit.  
In these regions, velocities are comparable to the speed of light and well known relativistic effects (such as gravitational redshift, 
light-bending, frame-dragging, and Doppler shifts) distort the iron line shape into a characteristic broad shape.
Measurements of the line profile then gives a detailed overview of the motion of matter very close to the central object.  
Among the available timing features, fast quasi periodic oscillations (QPOs) from BHs and NSs and coherent pulsations from the rotation 
of NSs and white dwarfs (WDs) are widely used to probe the intimate physical structure of these objects. 
QPOs in X-ray binaries and very massive BH in Active Galactic Nuclei (AGN), appear at the characteristic dynamical time-scales of the innermost 
disk regions, and are thus associated to the matter motion in strong field gravity. In the case of NSs, the detection and
accurate measurement of coherent pulsations has been demonstrated effective in constraining its mass and radius, thus providing information  
on the NS equation of state \citep{stella09,klis06,strohmayer06}.  

Many decisive progresses in all the above fields of research were made possible in particular through the exploitation 
of the time resolution and effective area of the proportional counters array (PCA, $\sim$0.67~m$^2$) 
on-board RXTE \citep{jahoda}. LOFT, the Large Observatory For X-ray Timing, will constitute the next giant leap forward with respect to any existing 
X-ray timing devoted experiment, and it is being proposed as the natural follow-up to the more than 15 years of operations of 
RXTE. The instruments on-board LOFT will be endowed with an unprecedentedly large effective area 
($\sim$10~m$^2$ at 8~keV) and a spectral resolution of $\sim$260~eV (FWHM at 6~keV). This will revolutionise the research field of compact 
objects and yield a great opportunity to address many 
of the fundamental questions on the matter under extreme conditions that were left open by the previous generation of X-ray instruments.   
\begin{figure}
\centering 
\includegraphics[scale=0.40]{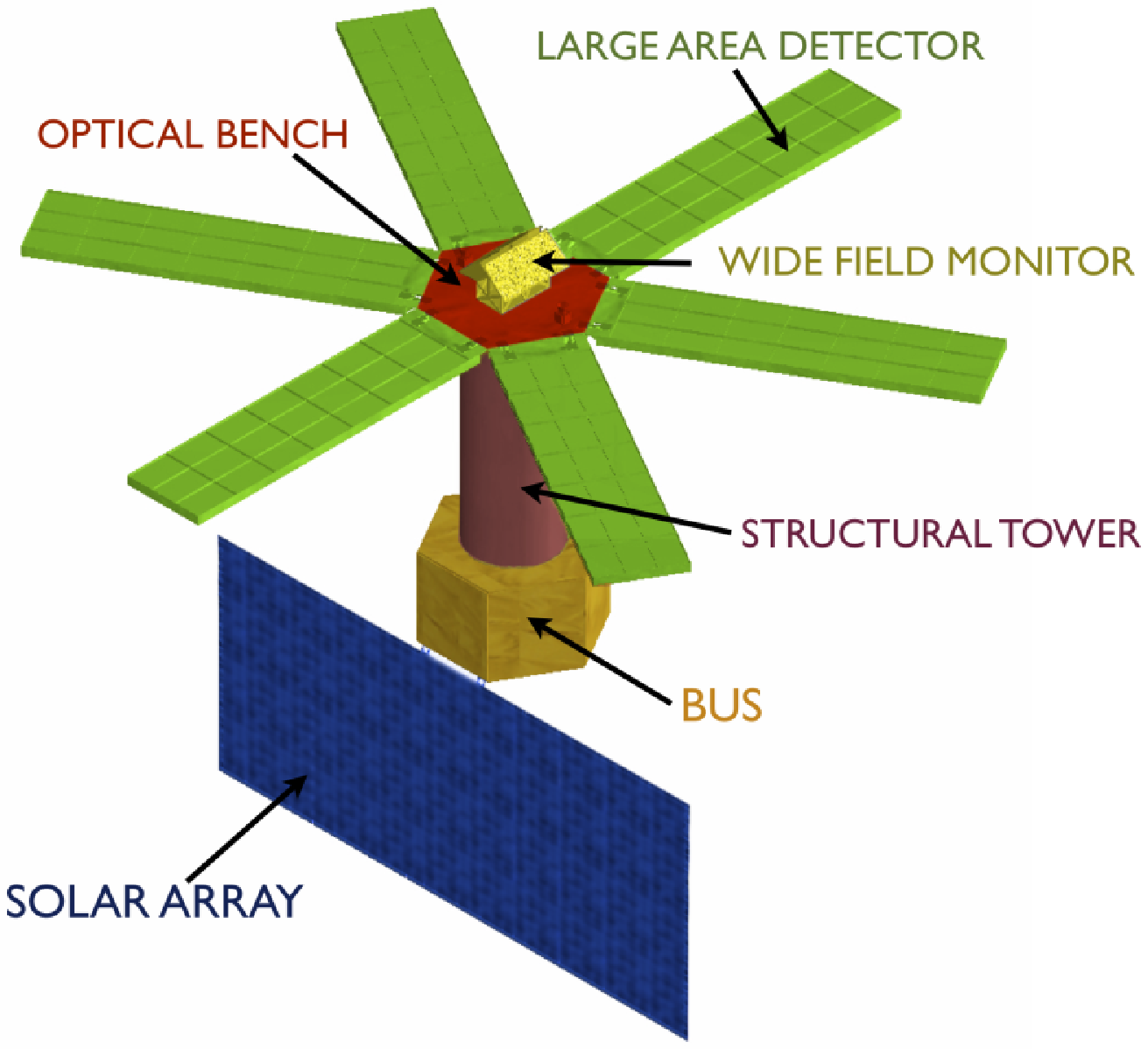}
\includegraphics[scale=0.32]{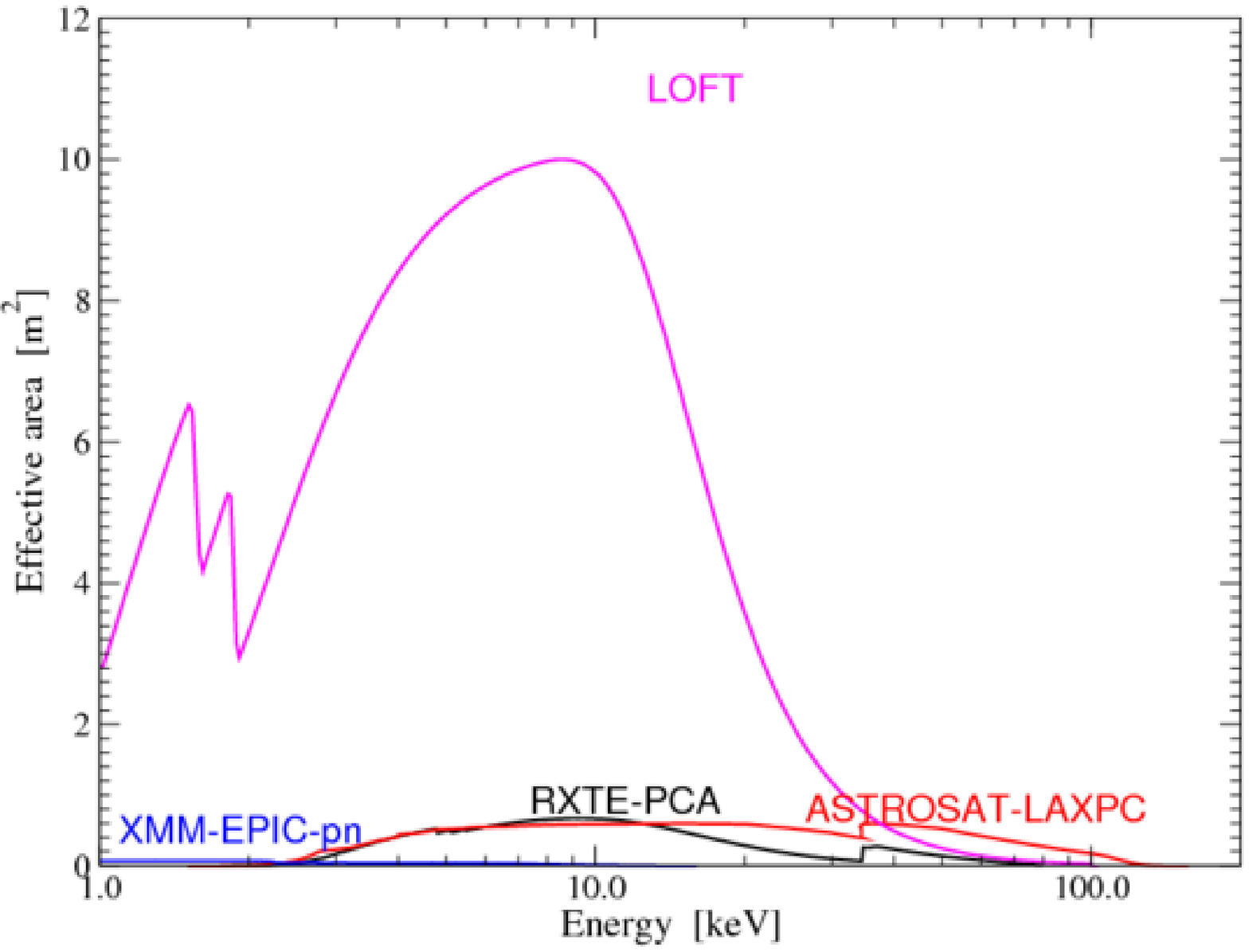}
\caption{\scriptsize {\it Left panel}: Sketch of the LOFT satellite. The 6 deployable panels of the LAD, the WFM and the solar array 
are indicated. {\it Right panel}: effective area of the LOFT/LAD as a function of energy.}
\label{fig:area} 
\end{figure}

\section{The concept}
\label{sec:concept} 

LOFT is specifically designed to exploit the diagnostics of rapid X-ray flux and spectral variability that directly 
probe the motion of matter down to distances very close to BHs and NSs. The successes achieved by 
RXTE in this context showed that a proper investigation of the energy-resolved timing properties of the X-ray emission from 
cosmic sources requires the accurate measurement of the time-of-arrival and energy of the largest  
number of photons from the target source \citep{jahoda}. The unambiguous identification of the target  
source in this type of experiment can be achieved by narrowing the field of view by means of an aperture 
collimator, down to a level large enough (typically $\simeq$1$^{\circ}$) to allow for pointing uncertainties 
yet small  enough to reduce the aperture background (cosmic diffuse X-ray background) 
and the risk of source confusion. 
In this type of instruments, the knowledge of the impact point of the photon on the detector array is not 
needed (if not for the use of proper detector calibration data), so position sensitive detectors are not required. 
Instead, detector read-out segmentation  is useful/necessary to reduce the effects of  pile-up and dead time.  

With these concepts in mind, the main instrument on-board LOFT \citep{feroci11} was designed as a collimated experiment, 
operating in the 2-30 keV energy range and reaching an effective area of $\sim$10~m$^2$ at 8~keV and a spectral resolution 
of $\sim$260~eV (FWHM) at 6~keV. This Large Area Detector (LAD) will provide 
a total of $\sim$140000 cts/s for a 500 mCrab source, thus constituting a real breakthrough with respect to 
any other existing X-ray instrument (see Fig.~\ref{fig:area}). 
The LOFT scientific payload  is completed by a coded-mask Wide Field Monitor (WFM) in charge of monitoring 
a large fraction of the sky potentially accessible to LAD, to provide the history and context  
for the sources observed by the LAD and trigger its observations on their most interesting states.  
The large field of view of the WFM will permit to observe in the same energy range of the LAD about 50\% of the sky at once. 
The WFM is designed also to catch transient/bursting events down to a few mCrab fluxes and will provide for them data with  
a spectral resoution of $<$500~eV in the 2-50 keV energy range and timing resolution of 10~$\mu$sec. 

The LOFT satellite is considered to operate in a low equatorial earth orbit ($\sim$600~km, $<$5$^{\circ}$~deg inclination) in order to 
reduce the background and the radiation damage effect of South Atlantic Anomaly. The science return of the mission has been so far 
evaluated assuming a medium-small class and a Vega launcher, even though other options are being considered\footnote{See also 
the results of the ESA Concurrent Design Facility (CDF) at http://sci.esa.int/science-e/www/object/index.cfm?fobjectid=49357}.

\subsection{The Large area detector}
\label{sec:lad} 

The Large Area Detector (LAD) of LOFT is designed as a classical collimated experiment. The key feature of the LAD design 
that allows reaching for the first time a very large effective area and a improved energy resolution is the low mass per unit 
area enabled by the solid-state detectors and capillary plate collimators (in the range of $\sim$10~kg~m$^{-2}$ compared to 
$>$100~kg~m$^{-2}$ of the RXTE/PCA). The basic set-up of the instrument is a set of 6 
Detector Panels tiled with $\sim$2000 Silicon Drift Detectors \citep[SDDs, heritage of the ALICE experiment 
at CERN/LHC;][]{vacchi91,zampa10,campana10}, which operate in 
the energy range 2-50~keV and have an energy resolution of $\sim$260~eV (FWHM, at 6~keV). The modular structure (see Fig.~\ref{fig:ladstructure}) 
ensures a high level of redundancy and the robustness of the instrument against single units failures. The field of view of the LAD 
is limited to $\sim$40 arcmin by X-ray collimators. These are developed by using the technique of micro-capillary plates, the 
same used for the micro-channel plates: a 3~mm thick sheet 
of Lead glass is perforated by a huge number of micro-pores, $\sim$20~$\mu$m diameter, $\sim$4-6~$\mu$m wall thickness. The stopping power of Lead  
in the glass over the large number of walls that off-axis photons need to cross is effective in collimating X-rays below 50~keV. 
In order to accommodate for the internal misalignments of the instrument and for attitude uncertainties, the response of the 
collimator in the central $\sim$10-15~arcmin angle is flat (flat-top response) to avoid any spurious modulation of the detected 
source flux. A summary of the presently established requirements and goals of the LAD are reported in Table~1. 
\begin{table}
\begin{minipage}[b]{0.45\linewidth}
\tiny
\begin{tabular}{@{}lll@{}}
\multicolumn{3}{c}{\bf LAD} \\
\hline
Parameter & Requirement & Goal \\
\hline
Energy range & 2--30~keV (nominal) & 1--30 keV (nominal) \\
             & 2--50~keV (expanded)& 1--50 keV (expanded) \\
\vspace{-0.2cm}\\
Eff. area & 10.0~m$^2$ (8~keV) & 12~m$^2$ (8~keV) \\
          & 1.0~m$^2$ (30 keV) & 1.2~m$^2$ (30 keV) \\
\vspace{-0.2cm}\\
$\Delta$E & $<$260~eV & $<$200~eV \\
(FWHM, @6 keV)  & (200~eV, 40\%)$^a$  & (160~eV, 40\%)$^a$ \\
\vspace{-0.2cm}\\
FoV (FWHM) & $<$60 arcmin & $<$30 arcmin \\
\vspace{-0.2cm}\\
Time res. & 10 $\mu$s & 7 $\mu$s \\
\vspace{-0.2cm}\\
Dead time & $<$1\% (@1 Crab) & $<$ 0.5\% (@1 Crab) \\
\vspace{-0.2cm}\\
Background flux & $<$10 mCrab & $<$ 5 mCrab \\
\vspace{-0.2cm}\\
Max. src flux (steady) & $>$0.5 Crab &  $>$0.5 Crab \\
\vspace{-0.2cm}\\
Max. src flux (peak) & $>$15 Crab &  $>$30 Crab \\
\hline
\multicolumn{3}{l}{$a$: Refers to single-anode events.} \\
\end{tabular}
\end{minipage}
\tiny
\centering
\hspace{0.5cm}
\begin{minipage}[b]{0.45\linewidth}
\begin{tabular}{@{}lll@{}}
\multicolumn{3}{c}{\bf WFM} \\
\hline
Parameter & Requirement & Goal \\
\hline
Energy range & 2--50~keV  & 1--50 keV  \\
\vspace{-0.2cm}\\
$\Delta$E (FWHM)& $<$500 eV & $<$300 eV \\
\vspace{-0.2cm}\\
FoV (FWHM) & 50\% of the sky & Same as requirements. \\
& accessible to & Sensitivity improvements  \\
& the LAD & are the prime goal\\
\vspace{-0.2cm}\\
Angular res. & 5 arcmin & 3 arcmin \\
\vspace{-0.2cm}\\
Position Accuracy& 1 arcmin & 0.5 arcmin \\
\vspace{-0.2cm}\\
Sensitivity & 5 mCrab & 2 mCrab \\
(5$\sigma$, 50 ks)  \\
\vspace{-0.2cm}\\
Sensitivity  & 1 Crab & 0.2 Crab \\
(5$\sigma$, 1 s) \\
\hline
\end{tabular}
\end{minipage}
\label{tab:lad}
\caption{\scriptsize {Scientific requirements and goals of the LOFT/LAD and LOFT/WFM.}}
\end{table}

\subsection{The wide field monitor}
\label{sec:wfm} 

\begin{figure}
\centering 
\includegraphics[scale=0.35]{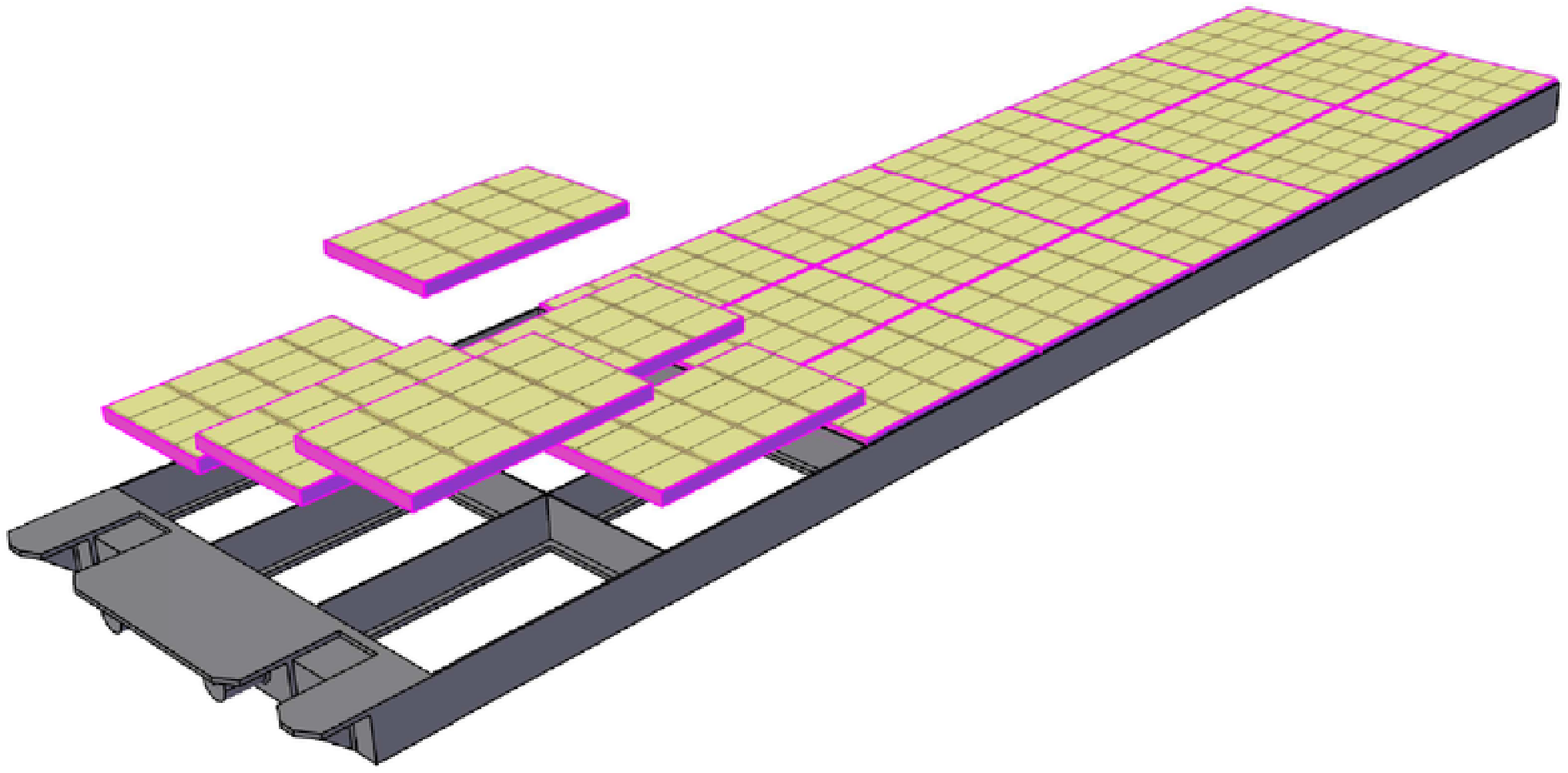}
\includegraphics[scale=0.35]{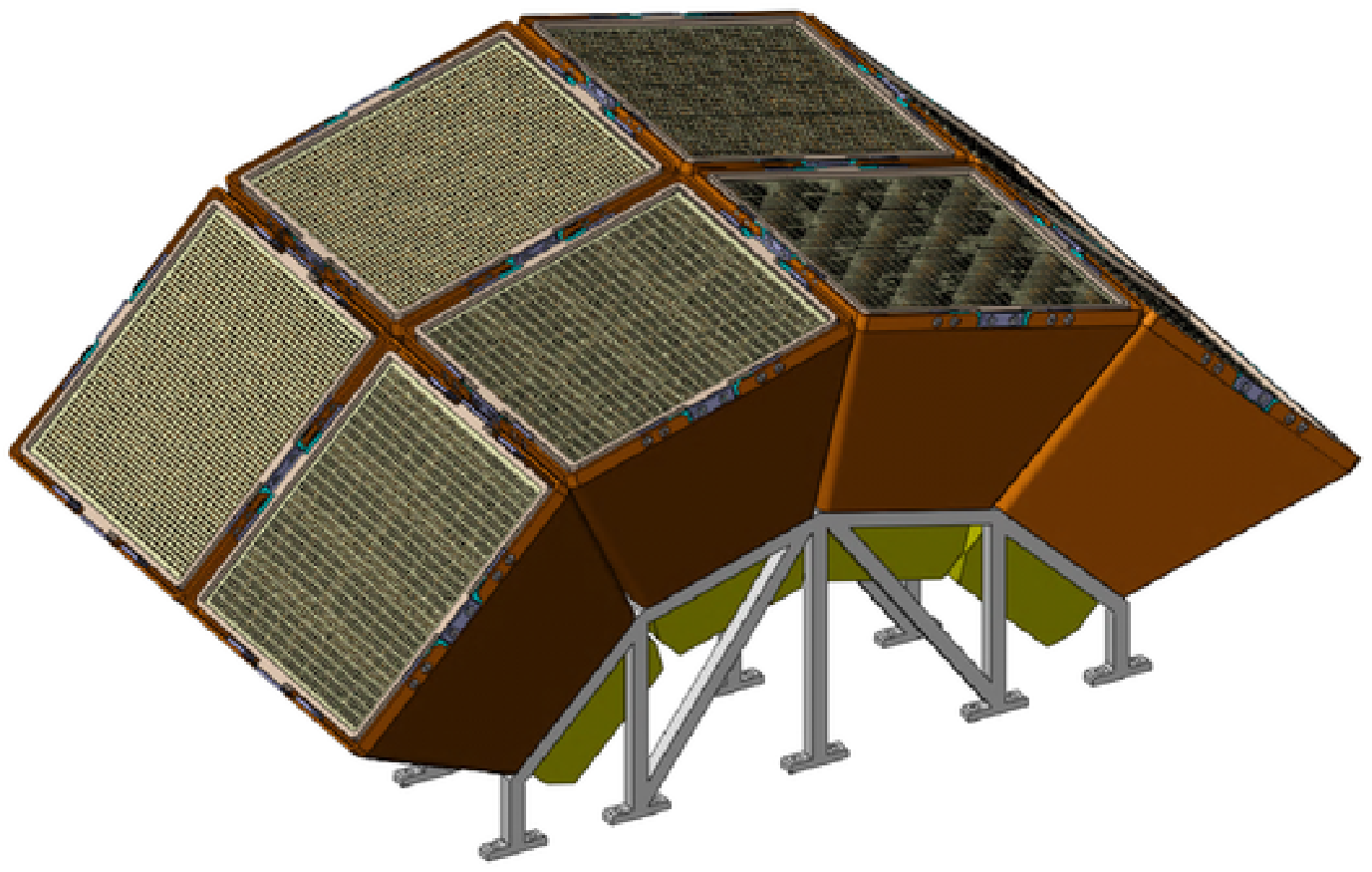}
\caption{\scriptsize {\it Left panel}: Details of the modular structure of one LAD panel (21 modules hosting 16 SDD each).  
{\it Right panel}: The current design of the WFM, comprising 4 units constituted by two co-aligned cameras each.}
\label{fig:ladstructure} 
\end{figure}

The LOFT baseline\footnote{An alternative solution is also under consideration which implies the usage of double sided Silicon strip detectors; see 
http://www.isdc.unige.ch/loft/index.php/instruments-on-board-loft.} 
WFM is a coded aperture imaging experiment designed on the heritage of the SuperAGILE experiment successfully 
operating in orbit since 2007 \citep{feroci07}. With the $\sim$100~$\mu$m position resolution provided by its Silicon microstrip detector, SuperAGILE 
demonstrated the feasibility of a compact, large-area, light, and low-power high resolution X-ray imager, with steradian-wide 
field of view. 
The LOFT WFM applies the same concept, with improvements provided by the higher performance (low energy threshold and energy 
resolution) of the Silicon Drift Detectors (SDDs) in place of the Si microstrips. The working principle of the WFM is the classical 
sky encoding by coded masks and is widely used in space borne instruments (e.g. INTEGRAL, RXTE/ASM, Swift/BAT). The mask shadow 
recorded by the position-sensitive detector can be deconvolved by using the proper procedures and recover the image of the sky, 
with an angular resolution given by the ratio between the mask element and the mask-detector distance.  By using SDDs with a 
position resolution $<$100~$\mu$m, a coded mask at $\sim$200~mm provides an angular resolution $<$5~arcmin. 
As a first approach, each WFM camera can be considered a one-dimensional coded mask imager. This means that after the proper 
deconvolution is applied to the detector images, the image of a sky region including a single point-like source will appear 
as a single peak over a flat background. The position of the peak corresponds to the projection of the sky coordinates onto 
the WFM reference frame. The width of the peak is the point spread function (few arc minutes for the LOFT/WFM). 
If more than one source is present in the observed sky region, the image will show a corresponding number of individual peaks, 
whose amplitude will depend on the intensity of the source and on the exposed detector area at that specific sky location. By 
observing simultaneously the same sky region with two cameras oriented at 90$^{\circ}$ to each other (such pair composing one WFM unit), 
one can derive the precise 2D position of the sources, by intersecting the two orthogonal 1D projections. The overall configuration of the 
WFM envisages at present a set of 4 units, each comprising 2 co-aligned cameras (see Fig.~\ref{fig:ladstructure}). 
The 4 units are off-set one to other along one direction in order to provide the maximum coverage of the region of the sky accessible to 
the LAD (see also Table~\ref{tab:lad}). 

\section{Conclusions}
\label{sec:conclusions} 
\begin{figure}
\centering 
\includegraphics[scale=0.22]{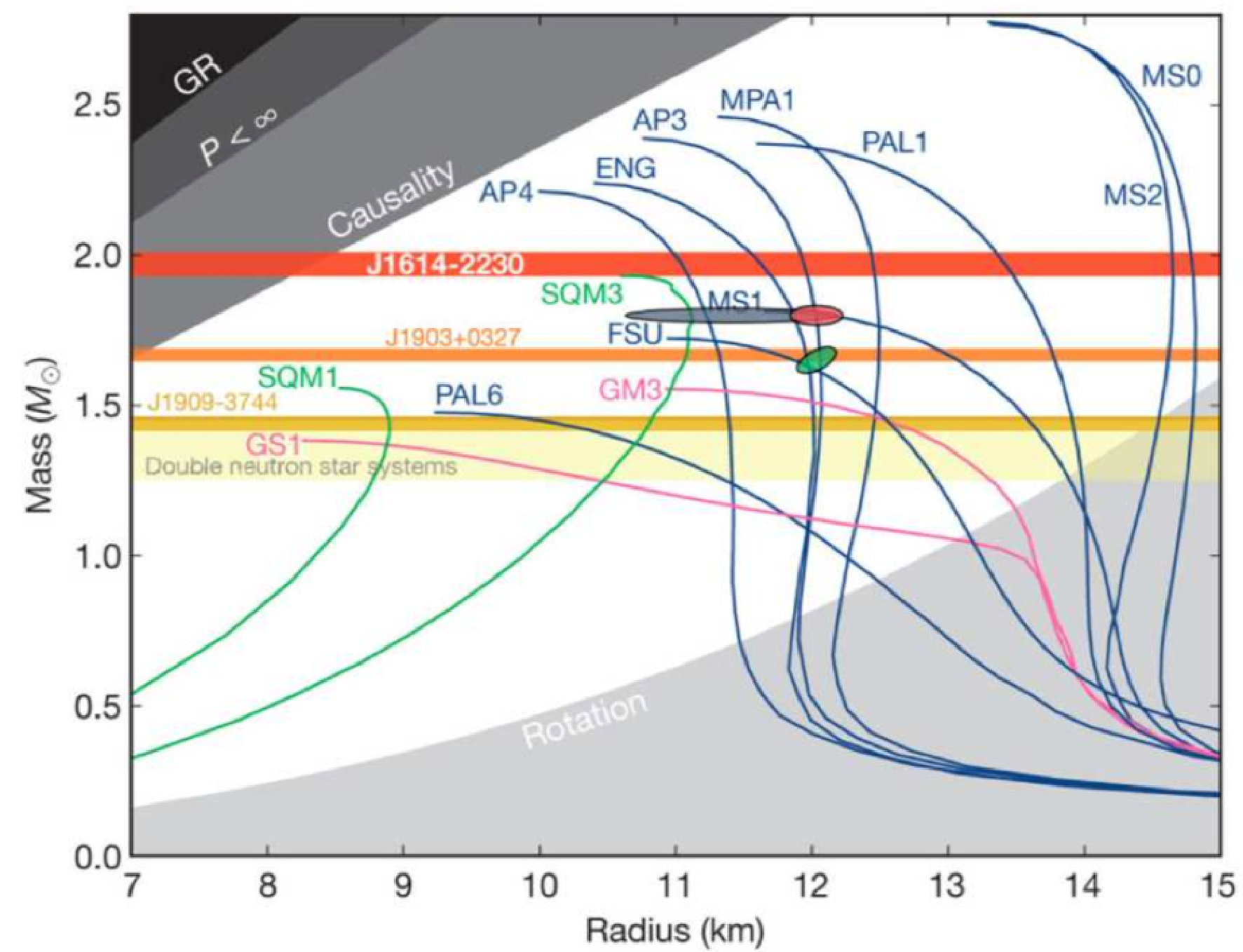}
\includegraphics[scale=0.30]{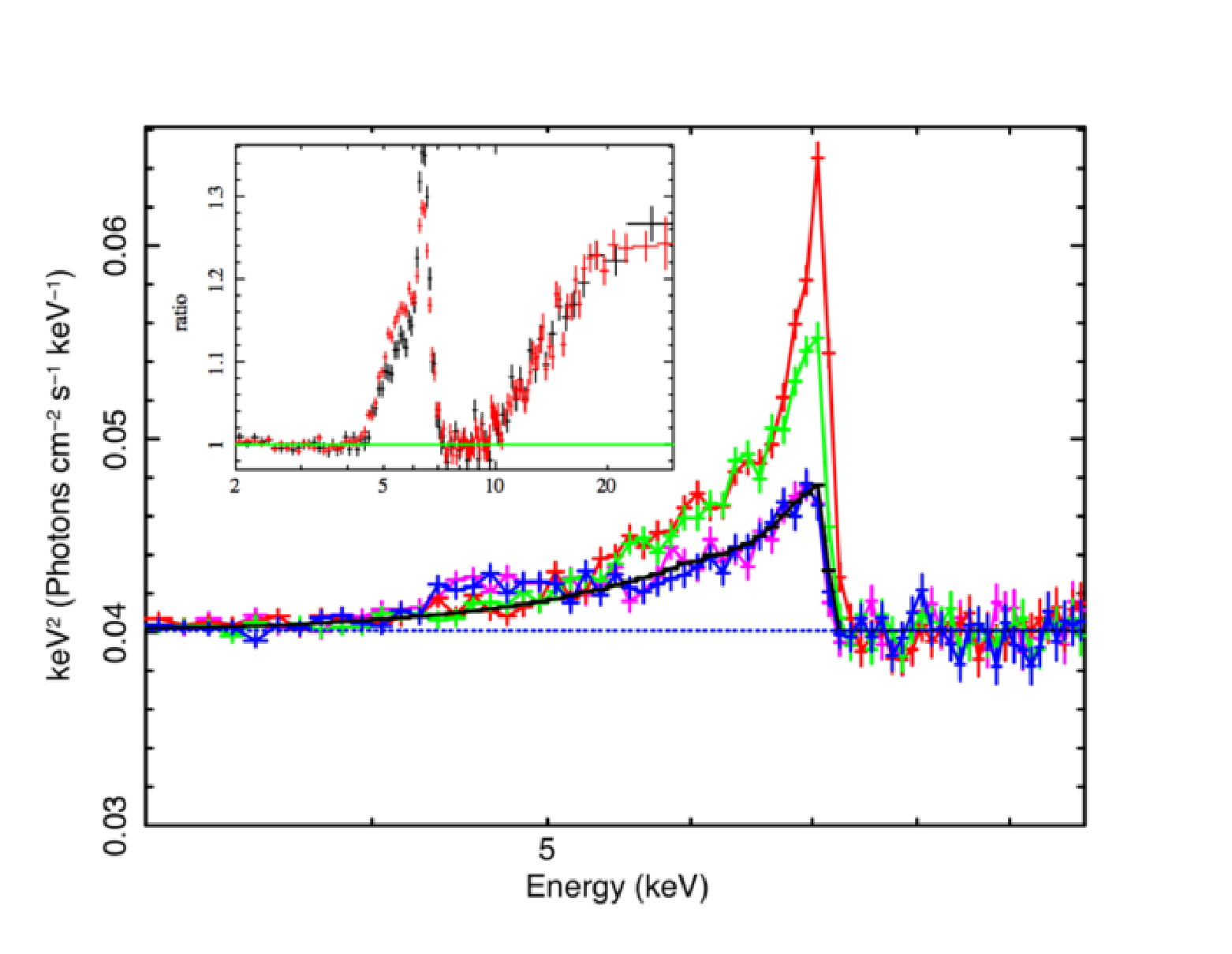}
\caption{\scriptsize {\it Left panel}: Mass-radius curves of representative EOSs for nucleonic (blue), nucleonic plus exotic (pink), and strange quark 
(green) matter. Horizontal bars represent mass determinations of binary radio pulsars. Regions to the upper left are disallowed by General 
Relativity and causality. The lower right region is excluded by the highest-frequency pulsar known (716 Hz). The pulse profile 
rising phase of a Type I X-ray burst was simulated by assuming the LAD capabilities in a way similar to \cite{strohmayer04}. 
The derived 90\% confidence level limits on the NS M and R from observations with LOFT are shown by the green ellipse. 
The uncertainty is $<$5\% on both M and R. The M-R constraint derived from fitting 
simulated pulse profiles of SAX\,J1808.4-3658, the first millisecond X-ray pulsar discovered (401 Hz), are shown by a gray ellipse. The same technique 
applied to simulations of the signal of the first eclipsing millisecond X-ray pulsars, Swift\,J1749.4-2807 (518 Hz), gives in a much smaller 
uncertainty region (red); this is mainly due to the higher system inclination. {\it Right panel}: Simulated LOFT/LAD spectra of a 3 mCrab 
(2-10 keV) AGN. The lower plot shows the broad relativistic Fe line produced in the innermost region of the accretion disc extending
down to the marginally stable for an almost-maximally rotating  BH ($a$=0.99) with mass of 3.6$\times$10$^6$~M${\odot}$, viewed at an 
inclination of 45$^{\circ}$. The plot shows the variable Fe K feature produced by an orbiting spot at $r$=10~$GM$/$c^2$ on the disk surface that is
illuminated by a flare. The orbital period is 4~ks and the total exposure is 16~ks. LOFT/LAD
will track the line variation on 1~ks time scale, thus allowing a determination of the orbital
period through measurements at four different orbital phases over four cycles. The inset
shows two 10~ks spectra from simulated high (6~mCrab in 2-10 keV) and low (2~mCrab)
flux states LOFT/LAD observations of MCG-6-30-15. Thanks to the wide energy band and
the large effective area above 10~keV, the variation of the reflection hump can be measured
with 2\% accuracy.}
\label{fig:science} 
\end{figure}
  
LOFT was primarily designed to address the fundamental questions raised
by the Cosmic Vision under the ``matter under extreme conditions'' theme,
through the study of the spectral and fast flux variability of compact X-ray
sources. The high throughput required by this type of investigation is achieved 
by an unprecedented large collecting area ($\sim$10~m$^2$ at 8~keV), coupled with a spectral resolution 
of $\sim$260~eV (FWHM at 6~keV) . 
A simulation of the LOFT capabilities is shown in Fig.~\ref{fig:science} for two of the 
main science drivers of the mission: the measurement of the NS mass and radius in Galactic X-ray binaries 
and the detection of broad iron lines in bright AGNs. The science investigations that will 
be accessible with LOFT are far too broad to be efficiently summarized here, and we remind 
the reader to the documentation available on the official LOFT website\footnote{http://www.isdc.unige.ch/loft}.  

Following the selection of LOFT as one of the four candidate M3 missions for the Cosmic Vision 
programme in 2011, a further down-selection is expected in 2013. This will bring only one of the candidate missions to 
a launch in the 2020s.


\begin{thebibliography}{99}

\bibitem[\protect\citeauthoryear{Campana et al.}{2011}]{campana10}
Campana, R. et al. 2011, NIMPA, 633, 22


\bibitem[\protect\citeauthoryear{Feroci et al.}{2011}]{feroci11}
Feroci, M. et al. 2011, ExA, in press (arXiv:1107.0436). 

\bibitem[\protect\citeauthoryear{Feroci et al.}{2007}]{feroci07}
Feroci, M. et al. 2007, NIMPA, 581, 728. 

\bibitem[\protect\citeauthoryear{Jahoda et al.}{1996}]{jahoda}
Jahoda, K. et al., SPIE 1996, 2808, 59. 

\bibitem[\protect\citeauthoryear{Psaltis et al.}{2008}]{psaltis08}
Psaltis, D., in Living Reviews in Relativity 2008, 11, 9. 

\bibitem[\protect\citeauthoryear{Stella et al.}{2009}]{stella09}
Stella, L., in Physics of Relativistic Objects in Compact Binaries, Springer Netherlands. 

\bibitem[\protect\citeauthoryear{Strohmayer}{2004}]{strohmayer04}
Strohmayer, E.T. 2004, AIPC, 714, 245. 

\bibitem[\protect\citeauthoryear{Strohmayer et al.}{2006}]{strohmayer06}
Strohmayer, E.T., Watts, A.L. 2006, ApJ, 653, 593.

\bibitem[\protect\citeauthoryear{Vacchi et al.}{1991}]{vacchi91}
Vacchi, A. et al. 1991, Nucl. Meth. Instr. A, 306, 187.

\bibitem[\protect\citeauthoryear{van der Klis et al.}{2006}]{klis06}
van Der Klis, M., in Compact Stellar X-ray Sources 2006, Cambridge University Press.

\bibitem[\protect\citeauthoryear{Zampa et al.}{2010}]{zampa10}
Zampa, G., et al. 2010, SPIE, 7732, 146 


\end{thebibliography}
\end{document}